\documentclass[3p,times,procedia]{elsarticle}
\usepackage{nupha_ecrc}
\newcommand {\sumc}     {{\tt super-MC}}
\newcommand {\trento}   {{\tt TRENTo}}
\newcommand {\hijing}   {{\tt HIJING}}
\newcommand {\urqmd}    {{\tt UrQMD}}
\newcommand {\vishnu}   {{\tt iEBE-VISHNU}}
\newcommand {\vish}     {{\tt VISHNU}}

\newcommand {\ipglasma} {IP-Glasma}

\volume{00}

\firstpage{1}

\journalname{Nuclear Physics A}

\runauth{}

\jid{nupha}

\jnltitlelogo{Nuclear Physics A}

\usepackage{amssymb}
\usepackage[figuresright]{rotating}

\begin{document}

\begin{frontmatter}



%% Instructions from Editor: Please use the following \dochead only in the preprint version (e-print arXiv etc.);
%% use empty \dochead{} when submitting to Nuclear Physics A!
\dochead{XXVIIIth International Conference on Ultrarelativistic Nucleus-Nucleus Collisions\\ (Quark Matter 2019)}
%% e.g. \dochead{17th International Conference on Dynamical Processes in Excited States of Solids}

\title{One fluid might not rule them all}


%\author[label1]{You Zhou}
%\ead{you.zhou@cern.ch}
%\author[label2]{Wenbin Zhao}
%\author[label2]{Huichao Song}
%\address[label1]{Niels Bohr Institute, University of Copenhagen, Blegdamsvej 17, 2100 Copenhagen, Denmark}
%\address[label2]{Department of Physics and State Key Laboratory of Nuclear Physics and Technology, Peking University, Beijing 100871, China}

\author[label1]{\underline{You Zhou}}
\ead{you.zhou@cern.ch}
\address[label1]{Niels Bohr Institute, University of Copenhagen, Blegdamsvej 17, 2100 Copenhagen, Denmark}
\author[label2,label3]{Wenbin Zhao}
\author[label2,label3]{Koichi Murase}
\author[label2,label3,label4]{Huichao Song}
\ead{Huichaosong@pku.edu.cn}
\address[label2]{Department of Physics and State Key Laboratory of Nuclear Physics and Technology, Peking University, Beijing 100871, China}
\address[label3]{Collaborative Innovation Center of Quantum Matter, Beijing 100871, China}
\address[label4]{Collaborative Innovation Center of Quantum Matter, Beijing 100871, China}

\begin{abstract}
In this proceeding, we present our recent investigations on hydrodynamic collectivity in high-multiplicity proton--proton collisions at $\sqrt{s}=$ 13 TeV using the \vishnu{} hybrid model with different initial condition models, called \hijing, \sumc{} and \trento. We find that with carefully tuned parameters, hydrodynamic simulations can give reasonable descriptions of the measured two-particle correlations. However, multi-particle single and mixed harmonics cumulants can not be described by hydrodynamics with these three initial conditions, even for the signs in a few cases. Further studies show that the non-linear response plays an important role in the hydrodynamic expansion of the p--p systems. Such an effect can change $c_2\{4\}$ from a negative value in the initial state to a positive value in the final state. The failure of the hydrodynamic description of multi-particle cumulant triggers the questions on whether the hydrodynamics can rule all collision systems, including p--p collisions at the LHC. 

\end{abstract}


\end{frontmatter}

% \linenumbers

\section{Introduction}
\label{section:introduction}

Ultra-relativistic heavy-ion collisions at the LHC provide a unique opportunity to study the Quark--Gluon Plasma (QGP). With the new developments on both flow measurements and model calculations, the understanding of the properties of the QGP and its fluctuating initial conditions have been improved to an unprecedented level. In addition to the detailed study of flow in Pb--Pb and Xe--Xe collisions, the flow phenomena in small collision systems like p--Pb and p--p collisions, which initially expected to serve as a reference data, have been studied in great detail~\cite{Dusling:2015gta,Song:2017wtw,Nagle:2018nvi}. Surprising observation of flow phenomena in these smaller collision systems has attracted a lot of attention. It challenges both the hydrodynamic model as the ``standard model'' in heavy-ion physics and the PYTHIA model as the ``standard tool'' for the minimum-bias p--p physics at the same time. While it has been commonly accepted that the observed large anisotropic flow is attributed to the creation of the QGP in heavy-ion collisions, the origin of the similar size of flow measured in small systems is less clear and currently under intense debate. In this proceeding, we will show the latest developments of hydrodynamic flow in p--p collisions, and discuss if hydrodynamics could be the right mechanism to describe the observed flow feature in experiments.

\section{Hydrodynamic model}
\label{section:model}

\begin{figure}[tbh]
\begin{center}
\includegraphics[width=0.9\textwidth]{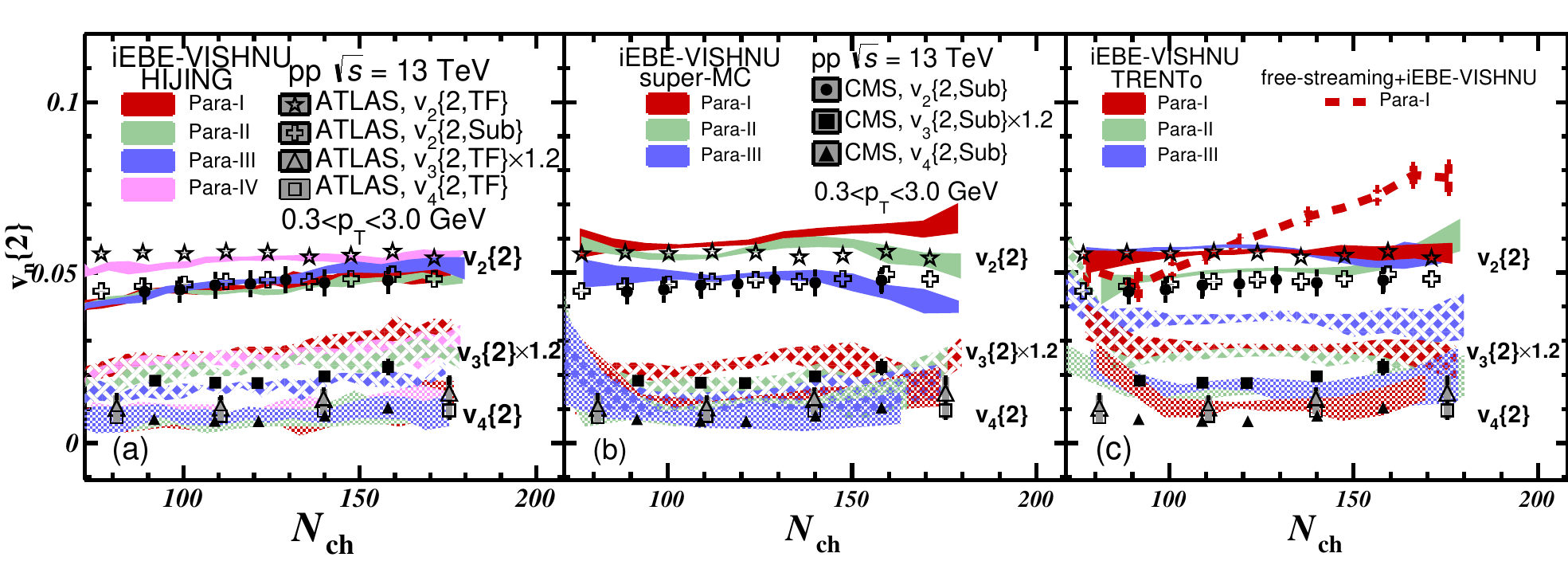}
\caption{
(Color online) Multiplicity dependence of $v_2$, $v_3$ and $v_4$ in p--p collisions at $\sqrt{s} = 13$ TeV\@, calculated by \vishnu{} with \hijing{} (left), \sumc{} (middle) and \trento{} (right) initial conditions, respectively. Para-I--IV represent different parameter sets tuned for each initial condition, which aretaken from Ref.~\cite{Zhao:2020pty}. 
}
\label{fig:fig1}
\end{center}
\end{figure}

In this study, \vishnu{} has been implemented to simulate the high multiplicity p--p collisions at $\sqrt{s}=$ 13 TeV. 
\vishnu~\cite{Shen:2014vra} is an event-by-event version of  \vish{} hybrid model that combines 2+1D viscous hydrodynamics and \urqmd{} hadron cascades to simulate the evolution of the QGP and hadronic matter. For simplicity, the bulk viscosity, net baryon density, and heat conductivity are neglected, and the specific shear viscosity $\eta/s$ is assumed to be constant. In order to investigate the dependence on the initial condition models, we implement three different initial condition models, namely, modified \hijing~\cite{Zhao:2017rgg}, \sumc~\cite{Welsh:2016siu}, and \trento~\cite{Moreland:2018gsh}. For more details please refer to~\cite{Zhao:2020pty}. It is found that with careful tuning of the input parameters, the \vishnu{} with each of the three different initial conditions is able to describe the measured integrated flow coefficients $v_2$, $v_3$ and $v_4$ with two-particle correlations as shown in Fig.~\ref{fig:fig1}.

\section{Results and discussions}
\label{section:results}

\begin{figure*}[thb]
\begin{center}
\includegraphics[width=0.9\textwidth]{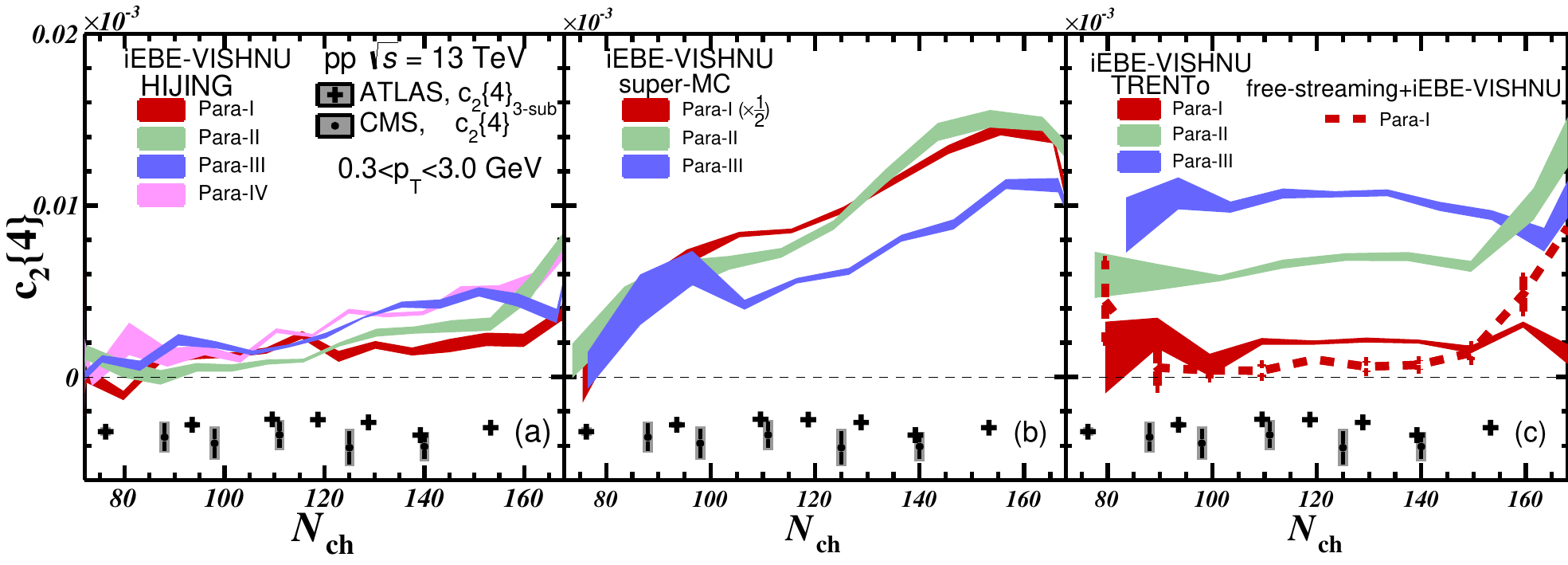}
\caption{(Color online)  $c_2\{4\}$ as a function of $N_{\rm ch}$ in p--p collisions at $\sqrt{s} =$ 13 TeV, calculated by  \vishnu{} with \hijing{} (left), \sumc{} (middle) and \trento{} (right) initial conditions, using standard cumulant method. The CMS data with standard cumulant method and the ATLAS data with three-subevent method are taken from~\cite{Khachatryan:2016txc} and~\cite{Aaboud:2017blb}, respectively.
}
\label{fig:fig2}
\end{center}
\end{figure*}

\begin{figure*}[t]
\begin{center}
\includegraphics[width=0.38\textwidth]{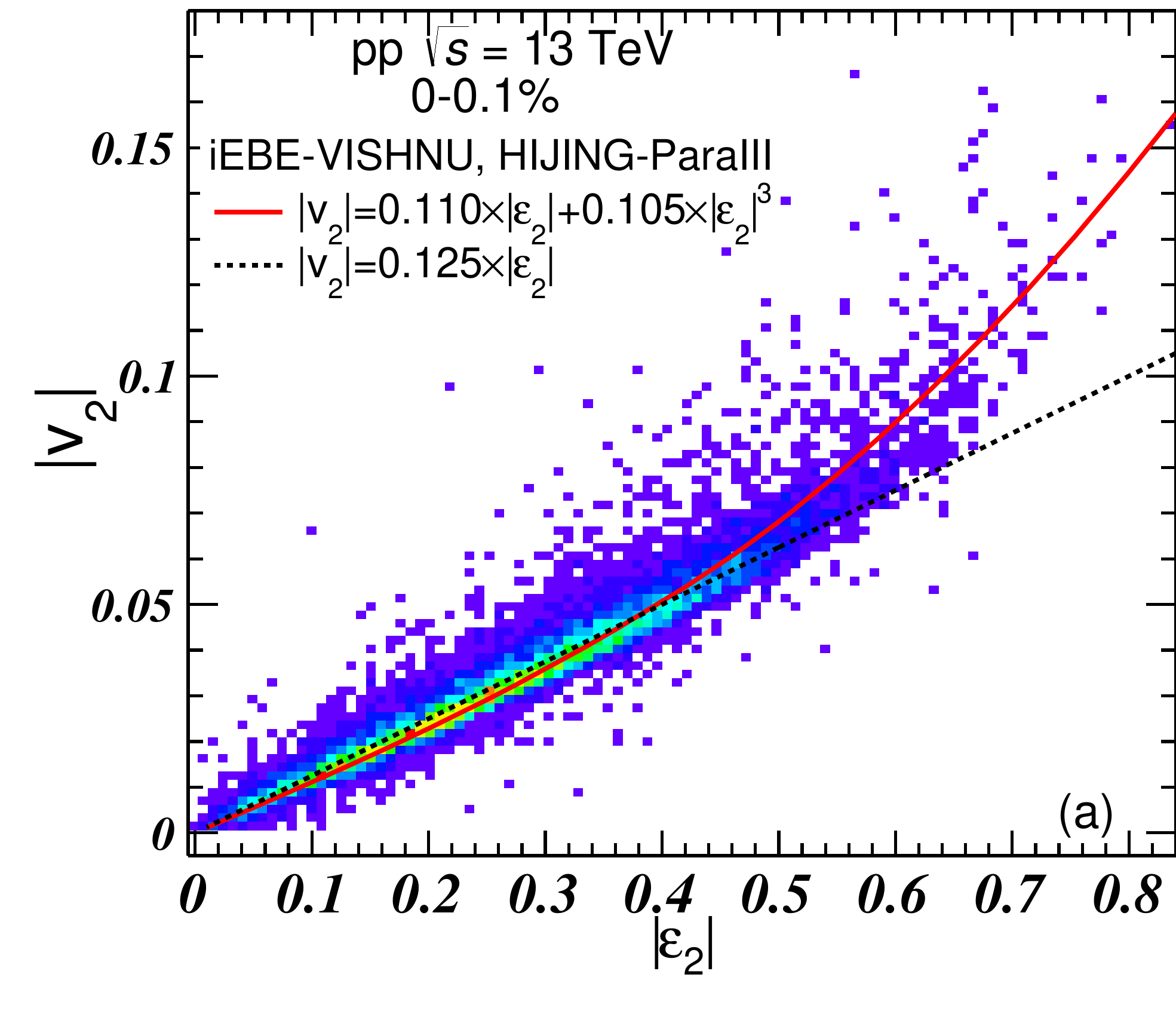}\includegraphics[width=0.38\textwidth]{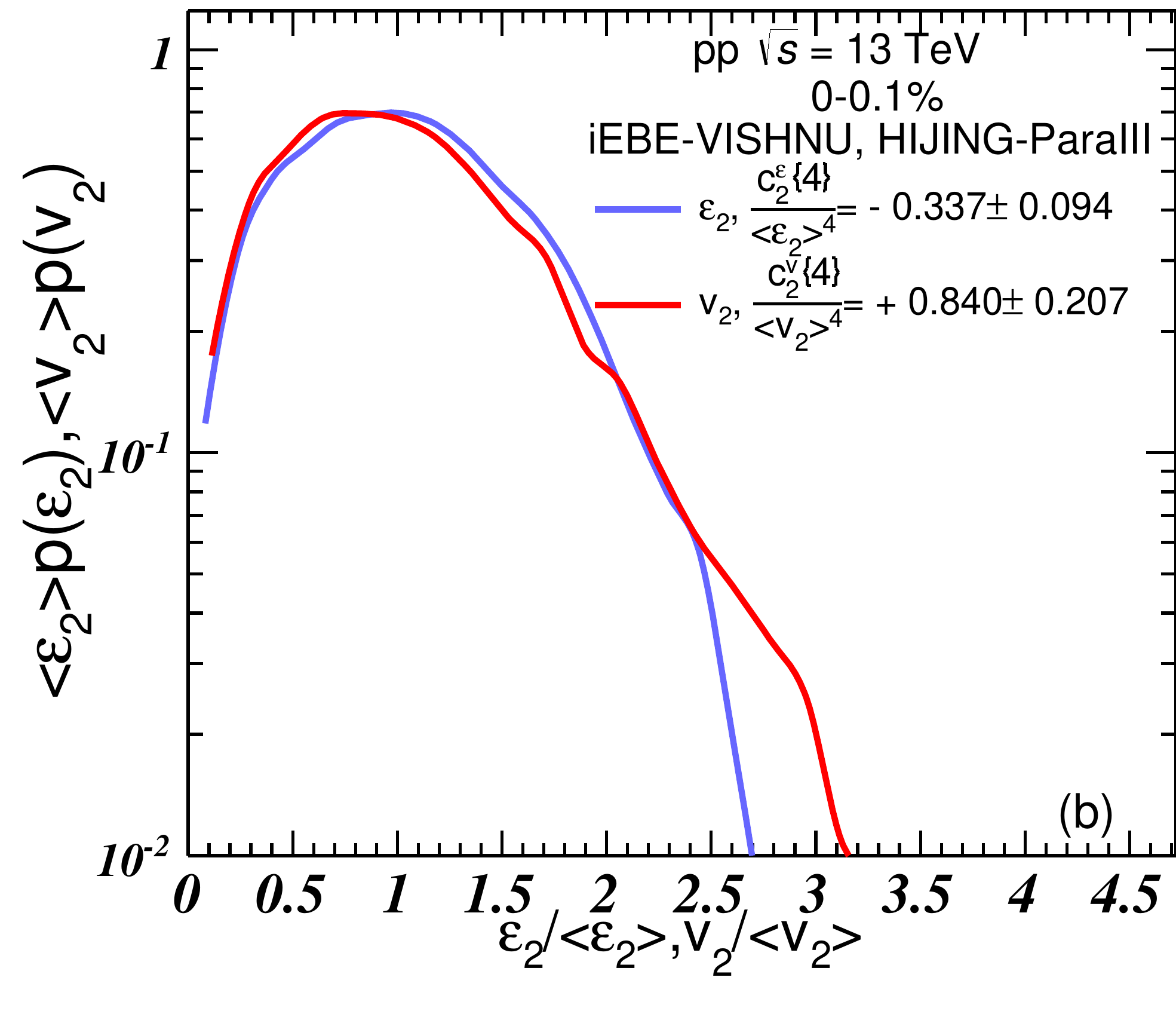}
\caption{(Color online) Left panel: the scatter points between the $v_2$ and $\varepsilon_2$ , together with a linear fitting and a non-linear fitting with cubic-term. Right panel: the comparison between the scaled event-by-event $\varepsilon_2$ distribution and scaled $v_2$ distributions of Para-III of \hijing{} initial conditions at 0--0.1\% centrality bin in p--p collisions at $\sqrt{s} =$ 13 TeV\@.  }
\label{fig:fig3}
\end{center}
\end{figure*}

%As shown above, within the \vishnu{} framework with \hijing, \sumc{} or \trento{} initial condition, we can qualitatively or even quantitatively describe the measured $v_n\{2\}$ (for $n=2$, $3$ and $4$) as a function of multiplicity. 
% The hydrodynamic framework fails to reproduce the measured negative $c_2\{4\}$ as in data. 
Figure~\ref{fig:fig2} shows the four-particle cumulants of the second harmonics, $c_2\{4\}$, in high multiplicity p--p collisions at $\sqrt{s} =$ 13 TeV\@. It shows that the hydrodynamic calculations of $c_2\{4\}$ are always positive, no matter which of the above-mentioned initial conditions has been applied. We also noticed that the hydrodynamic model MUSIC with \ipglasma{} initial conditions also do not generate a negative $c^v_2\{4\}$ in p--p collisions~\cite{Schenke:2020unx}. Thus, we emphasize that hydrodynamic expansion does not necessarily produce final state correlations with negative $c^v_2\{4\}$, and the observed negative $c^v_2\{4\}$ in experiments does not necessarily associate with the hydrodynamic flow in small collision systems~\cite{Zhao:2017rgg}.

In addition, we also find that the positive $c_2\{4\}$ from hydrodynamics is not due to the contaminations of non-flow or due to multiplicity fluctuations~\cite{Zhao:2017rgg}, but due to the large non-linear response of hydrodynamic evolution, as shown in the correlation between initial eccentricity $\varepsilon_2$ and final state $v_2$ in Fig.~\ref{fig:fig3} (left). A clear deviation of $v_2$ from the linear relationship with $\varepsilon_2$ is seen in the region of $\varepsilon_2 > 0.45$ , where a cubic response becomes significant~\cite{Teaney:2012ke, Noronha-Hostler:2015dbi}. The non-negligible non-linear response indicates that the scaled distribution of $v_2$, $P(v_2/\left<v_2\right>)$, doesn't follow the scaled distribution of $\varepsilon_2$, $P(\varepsilon_2/\left<\varepsilon_2\right>)$, which can be seen in Fig.~\ref{fig:fig3} (right). Such non-linear response leads to additional fluctuations of $v_2$ in the final states, which could even change the sign of $c_2^v\{4\}$ and eventually fail to reproduce the negative $c_2\{4\}$ as observed in experiment. 
Because two- and multi-particle cumulants have different sensitivities to flow fluctuations, the successful descriptions of two-particle correlations but failure descriptions of four-particle cumulants suggest that the presented hydrodynamic simulations in this proceeding could not describe the flow and the flow fluctuations simultaneously.

\begin{figure*}[thb]
\begin{center}
\includegraphics[width=0.9\textwidth]{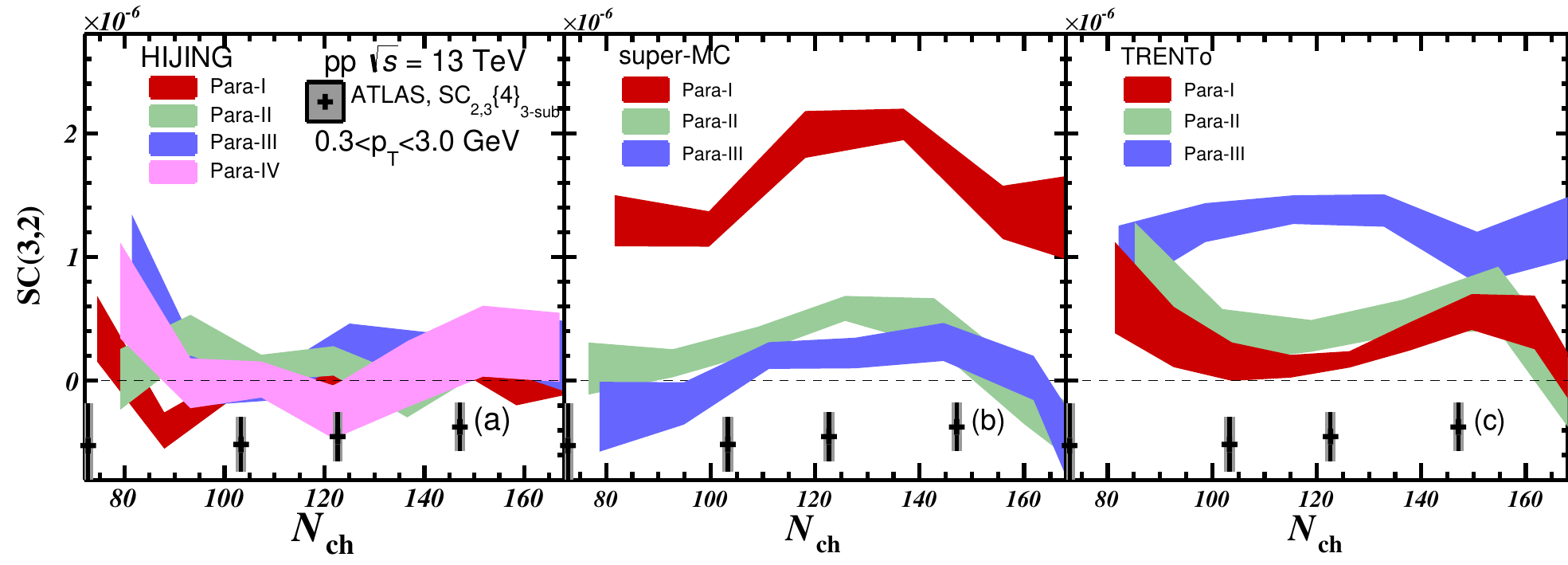}
\caption{(Color online)  Symmetric cumulant $SC(3,2)$ as a function of $N_{\rm ch}$ in p--p collisions at $\sqrt{s} =$ 13 TeV, calculated by  \vishnu{} with \hijing{} (left), \sumc{} (middle) and \trento{} (right) initial conditions using standard cumulant method.%The CMS data with standard cumulant method and the ATLAS data with three-subevent method are taken from~\cite{Khachatryan:2016txc} and~\cite{ATLAS:2017tqk}, respectively.
}
\label{fig:fig4}
\end{center}
\end{figure*}

Besides four-particle cumulants $c_2\{4\}$, we also study the multi-particle mixed harmonic cumulants in the high multiplicity p--p collisions at $\sqrt{s} =$ 13 TeV, using \vishnu{}  with different initial conditions. The three-particle asymmetric cumulant $ac_{2n,n,n} = \langle v_{n}^2 v_{2n} \cos 2n(\Psi_n-\Psi_{2n})\rangle $ is sensitive to the correlations between flow magnitudes and flow angles. The four-particle symmetric cumulants $SC(m,n)\ = \langle v_{m}^2v_n^2 \rangle-\langle v_{m}^2 \rangle\langle v_{n}^2 \rangle $ quantifies the correlation between $v_{m}^{2}$ and $v_{n}^{2}$~\cite{Bilandzic:2013kga}.
It was observed (but not shown in this proceedings) that the hydrodynamic calculations could qualitatively or even quantitatively describe $ac_n\{3\}$ and $nsc_{2,4}\{4\}$ after proper tuning of the parameters. Nevertheless, the results in Fig.~\ref{fig:fig4} show that the hydrodynamic calculations, independently on the initial conditions, could not describe the data. It seems that the currently applied hydrodynamic framework, which works nicely in the description of two-particle correlations, have some difficulties to describe the multi-particle single and mixed harmonic cumulants unless the implementation of the initial conditions are done with the assumption of the simple elliptic shape of the proton (when the proton is traveling in the near speed of light). In order to answer whether or not hydrodynamic flow has been produced in high multiplicity p--p collisions, further developments on initial conditions and the hydrodynamic modelling are both necessary.

\section{Summary}

In this proceeding, we reported our recent investigations on hydrodynamic flow in high-multiplicity proton--proton collisions at $\sqrt{s} =$ 13 TeV within the framework of \vishnu{} hybrid model with \hijing, \sumc{} and \trento{} initial conditions. We have shown successful descriptions of two-particle correlations and the current challenge to reproduce the multi-particle single and mixed harmonics cumulants. We leave the implementation of 3+1D hydrodynamics with dynamical initial conditions and longitudinal fluctuations to the future work to see if the hydrodynamic calculation has a chance to describe the existing fruitful data and thus confirm the creations of one tiny droplet of hydrodynamic fluid in p--p collisions at the LHC.

\section{Acknowledgments}
This work is supported by the Danish Council for Independent Research, the Danish National Research Foundation, the Carlsberg Foundation, a research grant (00025462) from VILLUM FONDEN, the NSFC under grant Nos.~11675004. We also gratefully acknowledge the extensive computing resources provided by the Super-computing Center of Chinese Academy of Science SCCAS, Tianhe-1A from the National Supercomputing Center in Tianjin, China and the High-performance Computing Platform of Peking University.



%% References


%\bibliographystyle{elsarticle-num}

\begin{thebibliography}{00}

%~\cite{Dusling:2015gta,Song:2017wtw,Nagle:2018nvi}

\bibitem{Dusling:2015gta}
  K.~Dusling, W.~Li and B.~Schenke,
  %``Novel collective phenomena in high-energy proton�Cproton and proton�Cnucleus collisions,''
  Int.\ J.\ Mod.\ Phys.\ E {\bf 25}, no. 01, 1630002 (2016).
 % doi:10.1142/S0218301316300022
  %[arXiv:1509.07939 [nucl-ex]].
  %%CITATION = doi:10.1142/S0218301316300022;%%
  %121 citations counted in INSPIRE as of 05 Oct 2019

%\cite{Song:2017wtw}
\bibitem{Song:2017wtw}
  H.~Song, Y.~Zhou and K.~Gajdosova,
  %``Collective flow and hydrodynamics in large and small systems at the LHC,''
  Nucl.\ Sci.\ Tech.\  {\bf 28}, no. 7, 99 (2017).
  %doi:10.1007/s41365-017-0245-4
 % [arXiv:1703.00670 [nucl-th]].
  %%CITATION = doi:10.1007/s41365-017-0245-4;%%
  %51 citations counted in INSPIRE as of 05 Oct 2019

\bibitem{Nagle:2018nvi}
  J.~L.~Nagle and W.~A.~Zajc,
  %``Small System Collectivity in Relativistic Hadronic and Nuclear Collisions,''
  Ann.\ Rev.\ Nucl.\ Part.\ Sci.\  {\bf 68}, 211 (2018).
  %doi:10.1146/annurev-nucl-101916-123209
  %[arXiv:1801.03477 [nucl-ex]].
  %%CITATION = doi:10.1146/annurev-nucl-101916-123209;%%
  %60 citations counted in INSPIRE as of 20 Jan 2020
  
  

    
\bibitem{Shen:2014vra}
  C.~Shen, Z.~Qiu, H.~Song, J.~Bernhard, S.~Bass and U.~Heinz,
  Comput.\ Phys.\ Commun. {\bf 199}, 61 (2016).
  
     \bibitem{Zhao:2017rgg}
  W.~Zhao, Y.~Zhou, H.~Xu, W.~Deng and H.~Song,
  %``Hydrodynamic collectivity in proton?proton collisions at 13 TeV,''
  Phys.\ Lett.\ B {\bf 780}, 495 (2018).
%  doi:10.1016/j.physletb.2018.03.022
%  [arXiv:1801.00271 [nucl-th]].
  %%CITATION = doi:10.1016/j.physletb.2018.03.022;%%
  %19 citations counted in INSPIRE as of 14 Jan 2020
  

%\cite{Welsh:2016siu}
\bibitem{Welsh:2016siu}
K.~Welsh, J.~Singer and U.~W.~Heinz,
%``Initial state fluctuations in collisions between light and heavy ions,''
Phys. Rev. C \textbf{94}, no.2, 024919 (2016)
% doi:10.1103/PhysRevC.94.024919
%[arXiv:1605.09418 [nucl-th]].
%42 citations counted in INSPIRE as of 28 Apr 2020

%\cite{Moreland:2018gsh}
\bibitem{Moreland:2018gsh}
J.~S.~Moreland, J.~E.~Bernhard and S.~A.~Bass,
%``Estimating initial state and quark-gluon plasma medium properties using a hybrid model with nucleon substructure calibrated to $p$-Pb and Pb-Pb collisions at $\sqrt{s_\mathrm{NN}}=5.02$ TeV,''
Phys. Rev. C \textbf{101}, no.2, 024911 (2020)
%doi:10.1103/PhysRevC.101.024911
% [arXiv:1808.02106 [nucl-th]].
%21 citations counted in INSPIRE as of 28 Apr 2020

%\cite{Zhao:2020pty}
\bibitem{Zhao:2020pty}
  W.~Zhao, Y.~Zhou, K.~Murase and H.~Song,
  %``Searching for small droplets of hydrodynamic fluid in proton--proton collisions at the LHC,''
  arXiv:2001.06742 [nucl-th].
  %%CITATION = ARXIV:2001.06742;%%

%\cite{Khachatryan:2016txc}
\bibitem{Khachatryan:2016txc}
  V.~Khachatryan {\it et al.} [CMS Collaboration],
  %``Evidence for collectivity in pp collisions at the LHC,''
  Phys.\ Lett.\ B {\bf 765}, 193 (2017).
  % doi:10.1016/j.physletb.2016.12.009
%  [arXiv:1606.06198 [nucl-ex]].
  %%CITATION = doi:10.1016/j.physletb.2016.12.009;%%
  %216 citations counted in INSPIRE as of 14 Jan 2020

%\cite{Aaboud:2017blb}
\bibitem{Aaboud:2017blb}
  M.~Aaboud {\it et al.} [ATLAS Collaboration],
  %``Measurement of long-range multiparticle azimuthal correlations with the subevent cumulant method in $pp$ and $p + Pb$ collisions with the ATLAS detector at the CERN Large Hadron Collider,''
  Phys.\ Rev.\ C {\bf 97}, 024904 (2018).
 % doi:10.1103/PhysRevC.97.024904
%  [arXiv:1708.03559 [hep-ex]].
  %%CITATION = doi:10.1103/PhysRevC.97.024904;%%
  %61 citations counted in INSPIRE as of 14 Jan 2020



 % \bibitem{BjoernQM}
%\cite{Schenke:2020unx}
\bibitem{Schenke:2020unx} 
  B.~Schenke, C.~Shen and P.~Tribedy,
  %``Bulk properties and multi-particle correlations in large and small systems,''
  arXiv:2001.09949 [nucl-th].
  %%CITATION = ARXIV:2001.09949;%%


%\cite{Teaney:2012ke}
\bibitem{Teaney:2012ke}
D.~Teaney and L.~Yan,
%``Non linearities in the harmonic spectrum of heavy ion collisions with ideal and viscous hydrodynamics,''
Phys. Rev. C \textbf{86}, 044908 (2012)
%doi:10.1103/PhysRevC.86.044908
%[arXiv:1206.1905 [nucl-th]].
%157 citations counted in INSPIRE as of 28 Apr 2020

%\cite{Noronha-Hostler:2015dbi}
\bibitem{Noronha-Hostler:2015dbi}
J.~Noronha-Hostler, L.~Yan, F.~G.~Gardim and J.~Ollitrault,
%``Linear and cubic response to the initial eccentricity in heavy-ion collisions,''
Phys. Rev. C \textbf{93}, no.1, 014909 (2016)
%doi:10.1103/PhysRevC.93.014909
% [arXiv:1511.03896 [nucl-th]].
%57 citations counted in INSPIRE as of 28 Apr 2020



%  \cite{Bilandzic:2013kga}
\bibitem{Bilandzic:2013kga}
 A.~Bilandzic, C.~H.~Christensen, K.~Gulbrandsen, A.~Hansen and Y.~Zhou,
  %``Generic framework for anisotropic flow analyses with multiparticle azimuthal correlations,''
    Phys.\ Rev.\ C {\bf 89}, 064904 (2014)
  % doi:10.1103/PhysRevC.89.064904
  % [arXiv:1312.3572 [nucl-ex]].
  %%CITATION = doi:10.1103/PhysRevC.89.064904;%%
  %134 citations counted in INSPIRE as of 14 Jan 2020




 \end{thebibliography}



\end{document}